# 1 kW 1 mJ 8-channel ultrafast fiber laser

MICHAEL MÜLLER,[1,*] MARCO KIENEL,[1,2] ARNO KLENKE,[1,2] THOMAS GOTTSCHALL,[1] EVGENY SHESTAEV,[1] MARCO PLÖTNER[3], JENS LIMPERT,[1,2,3] AND ANDREAS TÜNNERMANN[1,2,3]

[1]Friedrich-Schiller-University Jena, Abbe Center of Photonics, Institute of Applied Physics, Albert-Einstein-Straße 15, 07745 Jena, Germany
[2]Helmholtz-Institute Jena, Fröbelstieg 3, 07743 Jena, Germany
[3]Fraunhofer Institute for Applied Optics and Precision Engineering, Albert-Einstein-Straße 7, 07745 Jena, Germany
*Corresponding author: michael.mm.mueller@uni-jena.de

**An ultrafast fiber chirped-pulse amplifier comprising 8 coherently combined amplifier channels is presented. The laser delivers 1 kW average power at 1 mJ pulse energy and 260 fs pulse duration. Excellent beam quality and low-noise performance are confirmed. The laser has proven suitable for demanding scientific applications. Further power scaling is possible right away using even more amplifier channels.**

High-power ultrafast lasers are versatile tools for a wide range of industrial and scientific applications. In particular, Yb-doped slab [1], thin-disk [2], and fiber lasers [3] allow for very high average powers due to the low quantum defect heating and the superior heat dissipation. However, each laser geometry is facing technological and physical limitations that impede or even inhibit further power scaling. Regarding fiber lasers, currently these limitations are mode instabilities [4] and excessive nonlinearity. A successful technique to circumvent virtually any source specific limitation is coherent beam combination (CBC, [5]). An arbitrary number of laser beams is superposed and interferes constructively into a single beam given mutual coherence and a stable phase relation. Both conditions are fulfilled, e.g. when multiple laser amplifiers are seeded from a common source and when active phase stabilization is applied. Theoretically, the combined output brightness grows linearly with the number of beams combined, which in reality is not achievable as neither beam parameters nor pulse properties can be matched perfectly [6]. In this respect, a common figure of merit is the combining efficiency, which is defined as the ratio of the combined power to the sum of the individual beams' powers.

Although CBC can be applied to any laser geometry, fiber lasers are particularly suited for this approach as the modal guidance allows for very repeatable spatial, temporal, and spectral beam characteristics from different channels. Furthermore, the overall amplifier geometry is fairly simple, thus being robust and cost-effective. Recently, up to 530 W of average power at 1.3 mJ of pulse energy and even up to 5.7 mJ of pulse energy were demonstrated by coherent polarization beam combination of 4 large-mode-area fiber amplifiers [7,8]. In this letter, we present the newest advancement of this laser concept to an 8-amplifier-channel system, which allowed to increase the average output power of this ultrafast laser system to 1 kW. Thorough characterization of the output beam is presented for two points of operation. Power stability and further scaling potential are discussed towards the end.

Figure 1 schematically depicts the setup of the particular master-oscillator fiber-chirped-pulse-amplification system. The oscillator is an Yb:KGW Kerr-lens mode-locked oscillator delivering 60 fs pulses at 64 MHz repetition rate and 1 W average power. The pulse duration is increased to 1.3 ns FWHM duration in an Offner-type stretcher with a spectral hard cut of 22 nm centered at 1032 nm. The stretched pulses pass through a fiber-coupled spatial light modulator (SLM) to optimize pulse compression at the output.

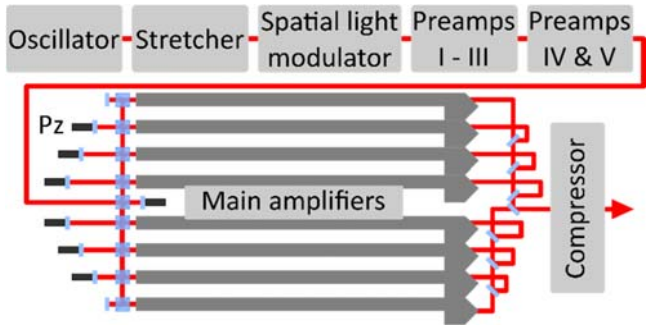

Fig. 1. Schematic of the 8-channel coherently-combined ultrafast fiber-laser system. Detailed description in text. (Pz: Piezo-driven mirror.)

The signal at this point has an average power of ~ 20 mW and is fed into an all-fiber section containing three 6 µm-core step-index fiber preamplifiers and two acousto-optic modulators in between those. This preamplifier block delivers an average power from 1 to 40 mW for repetition rates from 10 kHz to 64 MHz, respectively. After the all-fiber section, the beam is free-space coupled into a large-pitch fiber (LPF, [9]) preamplifier with 42 µm mode-field diameter and is amplified to an average power of 400 mW. Then, the signal is fed into a second LPF amplifier with 56 µm mode-field diameter making available a power of ~ 20 W for subsequent amplification. The seed beam is split up between the following 8 main amplifier channels using polarizing beam splitters and half-wave plates. First, the beam is split into two sub-beams from which power is channeled of consecutively into the actual fiber amplifiers. The beam division stages incorporate double-passes over piezo-driven mirrors mounted to manual translation stages. With the manual stages, the optical path length of all channel can be matched roughly. The piezo-driven mirrors allow to compensate fast phase fluctuations in a closed-loop control. The main amplifiers are LPFs with 64 µm mode-field diameter and lengths of 1.1m. Each fiber is pumped by a separate wavelength-stabilized fiber-coupled (200/240, 0.22 NA) pump diode delivering up to 250 W at 976 nm. The amplifiers work with circular polarization to minimize nonlinear phase accumulation [10]. After amplification, the beams are combined in free space using dielectric coated Brewster-type thin-film-polarizers (TFP) and half-wave plates (omitted in Fig. 1). Care is taken to match the beam parameters of each channel by fine tuning their collimation and by geometrically matching the optical path lengths in the combination stage. Laser windows directly after each TFP direct a small part of the signal to Hänsch-Couillaud polarization detectors [11] required for the active phase stabilization (omitted in Fig. 1). The beams are combined in the reverse order of the beam division. Thus, a straight link is given between the signal from each Hänsch-Couillaud detector and the associated piezo-driven mirror in each combination step. In total, 7 regulator channels are used to stabilize the relative phase between the 8 amplifier channels. Finally, the combined beam is enlarged to a diameter of about 7 mm and send through a Treacy-type compressor based on reflective dielectric gratings (1740 lines/mm) with a transmission efficiency of 90%.

The 8-channel amplifier stage was seeded with 17 W average power at 996 kHz repetition rate and was driven at maximum pump power. The compressed power from each channel was in the range of (138±7) W, which was combined to a compressed output power of 1003 W. This corresponds to a combination efficiency of 91% and pulse energy of 1 mJ. The autocorrelation of the combined output pulse had a duration of 368 fs (1.41*260 fs), as is depicted in Fig. 2. The B-integral from the amplifier chain is estimated to 1.5 rad assuming exponential amplification, which is easily compensable with phase shaping.

Figure 3 shows the combined output spectrum with an FWHM bandwidth of 5.8 nm. The spectral hard-cut of the stretcher/compressor is visible only on the long-wavelength end of the spectrum, slightly beyond 1040 nm. The dip at 1025 nm and the overall spectral modulation is introduced by the pixel mask of the SLM. The SLM's 128 pixels are spanning 25 nm bandwidth, which yields the modulation period of ~ 0.2 nm. Given the low nonlinearity, the impact of the modulation on the pulse quality is negligible.

The compressed beam profile of the individual channels and the beam profile of the coherently combined are depicted in Fig. 4. All beams are slightly elliptical after transmission through the compressor. Slight variations in size of the individual beams and differences in orientation of their underlying hexagonal structure (LPF's symmetry) vanish upon combination and result in a clean Gaussian beam.

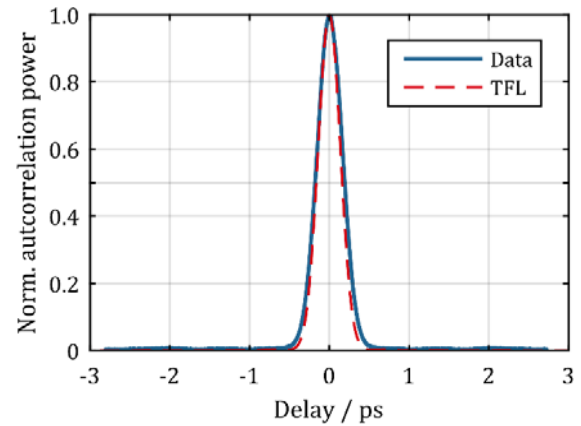

Fig. 2. Measured autocorrelation (Data) and transform-limited autocorrelation (TFL) at 1 mJ pulse energy and 1 kW average power.

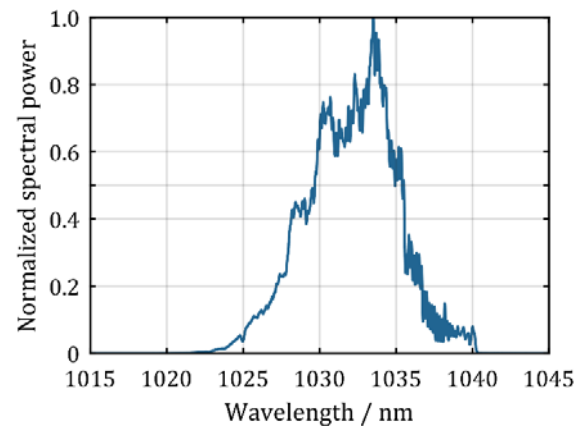

Fig. 3. Optical spectrum of the output at 1 mJ pulse energy and 1 kW average power.

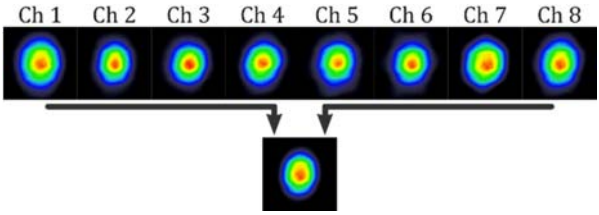

Fig. 4. Far-field beam profiles of the individual channels and after coherent combination of all channels at 1 kW of output power

The beam quality was analyzed by repeated $M^2$-measurements using the 4σ-method showing an almost diffraction limited beam quality of $M^2 < 1.1$ on both axes, as is shown in Fig. 5. The slight astigmatism did not degrade the beam quality. Hence, it can be compensated for by introducing the opposite sign aberration.

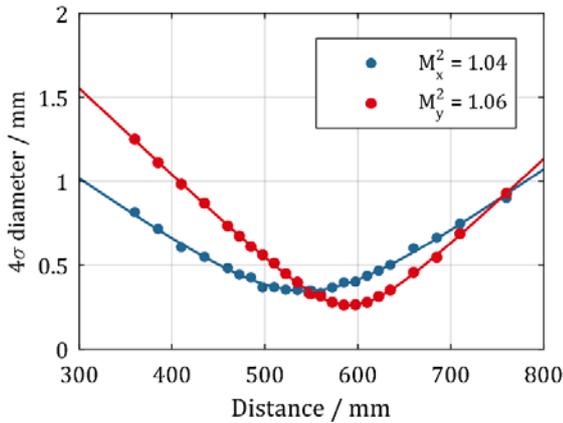

Fig. 5. $M^2$-measurement (4σ-method) of the 1 kW coherently combined beam parallel (blue) and vertical (red) in respect to the optical table.

The laser system was pushed towards higher pulse energy at a reduced repetition rate of 265 kHz. An average power of 874 W was achieved, corresponding to a pulse energy of 3.3 mJ. The average power of the individual channels was in the range of (123±7) W, which gives a combining efficiency of 89%. Due to the increased nonlinearity at this pulse energy, nonlinear polarization rotation was observed in the channels. This effect was compensated for by adjusting quarter- and half-wave plates in front of each channel and by slight variation of the individual pump powers. However, the compensation was not perfect, causing the slight decrease in combination efficiency. The B-integral in this case is estimated to be 4.7 rad, still assuming an exponential amplification. This high level of accumulated nonlinear phase explains the increased width of the autocorrelation to 388 fs (1.41*275 fs) despite phase shaping, as depicted in Fig. 6.

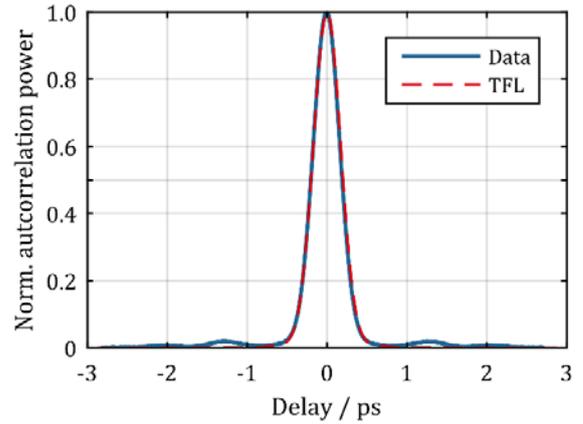

Fig. 6. Measured autocorrelation (Data) and transform-limited autocorrelation (TFL) at 3.3 mJ pulse energy showing the onset of uncompensated nonlinear phase.

Figure 7 shows the corresponding output spectrum with a reduced FWHM bandwidth of 5.0 nm. Compared to the 1 mJ-case, this reduction is due to both gain narrowing and the nonlinear polarization rotation acting as a spectral filter in the CPA regime. The nonlinearity also intensified the fine spectral modulation [12].

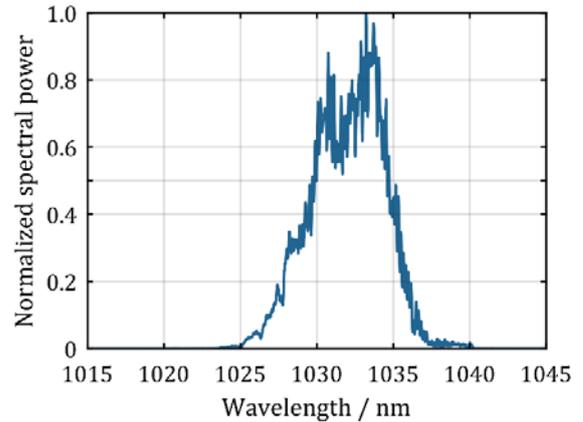

Fig. 7. Optical spectrum of the 3.3 mJ pulses at 870 W average power.

The combined beam held the almost diffraction limited quality with an $M^2$ value of 1.1 on both axes, as shown in Fig. 8. The beam sample for this measurement was taken closely to the compressor output, because a degradation of beam quality was found when taking the beam sample after propagating the high-power beam for about 4 m after compression. In that case the $M^2$ increased to 1.7, which is attributed to nonlinear propagation at the onset of ionization in air given the small beam diameter of ~7mm [13]. Aiding this interpretation, the beam quality was recovered when detuning the compressed pulse duration to ~2 ps. Hence, a larger beam diameter must be chosen when delivering the beam over a certain distance to an actual experiment.

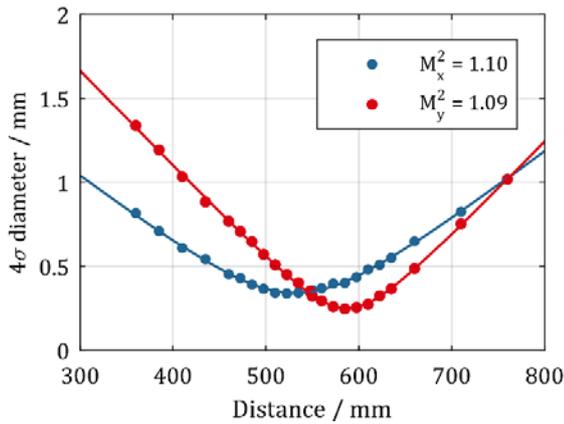

Fig. 8. $M^2$-measurement ($4\sigma$-method) of the combined beam at 3.3 mJ pulse energy parallel (blue) and vertical (red) to the optical table.

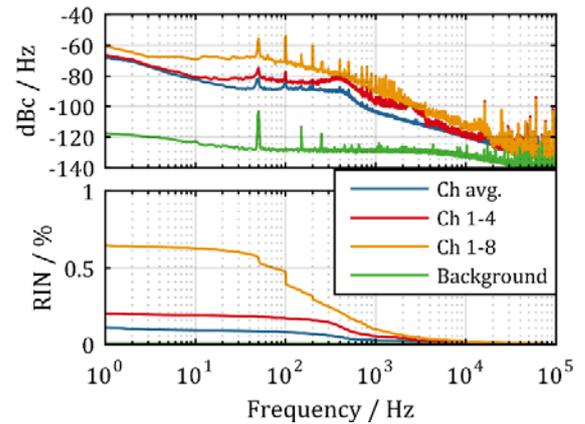

Fig. 9. Power spectral density (top) and relative intensity noise (bottom) at different positions in the laser system. The equivalent noise bandwidth of the Hanning window used in this analysis is 1.5 Hz.

The power stability of the 1 kW combined laser beam was analyzed in the base band using an amplified photo-diode (Thorlabs PDA36A-EC, 20dB gain) connected to a digital sampling scope (LeCroy HDO4096) through a 500 kHz low-pass filter (Thorlabs EF506). The laser repetition rate was set to 2.1 MHz in order to achieve sufficient suppression of the fundamental repetition rate in the RF-signal at the particular filter bandwidth. Thus, aliasing is avoided at the given sampling rate of 2.5 MSp/s. The beam size was much larger than the detector area, which implies that both pointing and amplitude fluctuations were detected. The top section of Fig. 9 depicts the resultant power spectral density for an average single channel, the coherent combination of 4 channels and the combination of all 8 channels as well as the measurement background in the frequency range from 1 Hz to 100 kHz. The bottom part of the figure depicts the integrated power spectral densities, with the relative intensity noise (RIN) being its maximum. The measurement background noise corresponds to a RIN of 0.006%. On average, the RIN for the compressed output of an individual main amplifier at an average power level of (138±7) W is 0.11%. Coherent combination of 4 channels results in a slight increase of the RIN to 0.20%. When all 8 channels were combined the RIN increased further to 0.65%, which still is a remarkably low value. The increase is attributed mostly to the significantly higher heat load when all channels are combined, which can be solved by proper engineering. Currently, phase and pointing fluctuations are not measured separately, but is subject to future experimental investigation.

Finally, to confirm applicability of the laser system, two experiments were performed at a moderate average power level of 600 W. First, the laser output was frequency doubled and tripled in bonded Sapphire-BBO-Sapphire stacks, which allowed to generate sub-ps pulses at center wavelength of 343 nm with an average power of 100 W [14]. The second experiment was a two-stage nonlinear compression in gas-filled hollow-core capillaries where sub-2-cycle pulses were generated at 200 W of average power [15].

Regarding further power scaling, the laser system currently is limited by the number of amplifier channels and the pump power available. Furthermore, heating of the optical components was observed using a thermal camera, but, as the $M^2$-measurements indicate, no degradation in beam quality was observed yet. Hence, further scaling is possible by using more amplifier channels at least until thermal lensing becomes an issue in the combination stage. In that case, improved optical components will be required.

In summary, a high-power ultrafast fiber-CPA system is presented, that incorporates coherent beam combination of 8 amplifier channels. The system delivers 1 kW average power at 1 mJ pulse energy. Moreover, 870 W average power at 3.3 mJ pulse energy are demonstrated. Always, the pulse duration is equal or less than 275 fs and the $M^2$-value is equal or less than 1.1 on both axes. The relative intensity noise is 0.65% in the frequency span from 1 Hz to 100 kHz. Two experiments were performed with this laser demonstrating its applicability to scientific research. The thermal load on the optical components is identified as the most probable future scaling limitation, which can be pushed out by higher-quality optics. Currently, a 16 channel laser system is being developed which is expected to deliver up to 2 kW of average power and pulse energies of 10 mJ.

**Funding.** This work has been partly supported by German Federal Ministry of Education and Research (BMBF) under contract 13N13167 'MEDUSA', the European Research Council (ERC): Grants 617173 'ACOPS' and 670557 'MIMAS'.

**Acknowledgment**. M.M. acknowledges financial support by the Carl-Zeiss-Stiftung. M.K. acknowledges financial support by the Helmholtz-Institute Jena. All authors would like to thank Active Fiber Systems GmbH for providing a pump diode.